\documentstyle[12pt,aasms4,epsf]{article}

\def\S{X0115+63}   \def\B{{\em BeppoSAX}}
\makeatletter
\def\preprint{preprint}   \newif\ifPreprintMode
\ifx\preprint\revtex@genre\PreprintModetrue\else\PreprintModefalse\fi
\makeatother

\lefthead{A. Santangelo et~al.}
\righthead{Multiple cyclotron harmonics in \S}

\begin{document}

\title{A {\it BeppoSAX} study of the pulsating transient \S : \\
the first X-ray spectrum with four cyclotron harmonic features}

\author{A. Santangelo, A. Segreto and S. Giarrusso}
\affil{Istituto Fisica Cosmica e Applicazioni all'Informatica (IFCAI), C.N.R.
\\
Via La Malfa 153, 90146 Palermo, Italy}

\author{D. Dal~Fiume and M. Orlandini}
\affil{Istituto Tecnologie e Studio Radiazioni Extraterrestri (TeSRE), C.N.R. \\
Via Gobetti 101, 40129 Bologna, Italy}

\author{A. N. Parmar and T. Oosterbroek }
\affil{Astrophysics Division, Space Science Department of ESA, ESTEC, \\
Keplerlaan 1, 2200 AG Noordwijk, The Netherlands}

\author{T. Bulik}
\affil{Nicolaus Copernicus Astronomical Center \\
Bartcyka 18, 00716 Warsaw, Poland}

\author{T. Mihara}
\affil{Institute of Physical and Chemical Research (RIKEN), \\
2--1, Hirosawa, Wako, Saitama, 351--0198, Japan}

\author{S. Campana}
\affil{Osservatorio Astronomico di Brera, \\
Via Bianchi 46, 23807 Merate (Lc), Italy}

\author{G.L. Israel and L. Stella }
\affil{Osservatorio Astronomico di Roma, \\
Via Frascati 33, 00040 Monteporzio Catone (Roma), Italy}

\begin{abstract}

The recurrent hard pulsating X--ray transient \S\ was
observed with \B\ on March 19, when the
source was at a 2--10~keV flux level of $\sim$310~mCrab. 
We report on the high energy spectrum of the source, concentrating on
cyclotron resonant scattering features. The spectrum is strongly pulse phase
dependent and absorption features are detected at virtually all phases. In
particular, four absorption-like features at 12.74~keV, 24.16~keV, 35.74~keV, 
and 49.5~keV
are observed in the descending edge of the main peak of the pulse profile.
The ratios between the centroid energies of the lines with respect to the first 
are 1:(1.9):(2.8):(3.9). 
These values are close to the
harmonic relation expected from cyclotron resonant scattering in a strong 
magnetic field, when relativistic effects are taken into account. 
The equivalent widths of the
second, third, and fourth harmonics are found to be larger than that of the
first harmonic, confirming the key role of two-photon processes in the 
spectral formation.
These results provide the first evidence for {\it four harmonically spaced
lines} in the spectrum of an accreting X-ray pulsar, yielding 
the clearest confirmation to date of their magnetic origin.

\end{abstract}

\keywords{binaries: close --- stars: neutron --- pulsars: individual: \S\ ---
          X--rays: stars} 

\section{Introduction}

Cyclotron features provide a powerful tool 
for directly measuring 
the high ($B\ge 10^{12}$~G) magnetic field strengths
of accreting neutron stars in X-ray binaries. 
Because the electron cyclotron energy is 
$E_{\rm cyc}=11.6B_{12}$~keV, where $B_{12}$ is the magnetic field
strength in unit of $10^{12}$~G, these features are expected to be observed at 
hard X--ray energies.
Absorption-like features interpreted as cyclotron resonant 
scattering were first discovered in the spectrum of the low-mass X-ray binary 
pulsar Her X--1 (\cite{Tru1978}) and, subsequently, in the hard X--ray 
transient pulsar \S\ (\cite{Whe79}). Since then, cyclotron features have been 
detected in other X--ray binary pulsars with {\it Ginga} 
(\cite{Miha95}), HEXE/TTM on Mir (\cite{kendz94}), OSSE onboard CGRO 
(\cite{Grove95})
and, more recently, with {\it 
RXTE} (\cite{kre98}) and \B\ (\cite{DAL99,Sant99}). 

Relatively little is known on higher cyclotron harmonics. 
Besides the pioneering detection of two lines in the spectrum 
of \S\ (\cite{Whi83}), the presence of two cyclotron lines 
has been reported for 
Vela X--1 (\cite{kre98,orl98}), 4U1907+09
(\cite{Cus98,Sant99}) and A0535+26 (\cite{kendz94}). 
Some of these detections are still to be confirmed. 
   
\S\ is one of the best studied X--ray transients
(\cite{rap78}). The source shows pulsation at $\sim 3.6$~s while orbiting an 
O9e companion (V635 Cassiopeiae, \cite{Unger98}) with a period of 24.3 days. 
As customary for this class of X-ray binaries, the X-ray continuum  
has been modelled with 
a power-law with an exponential 
cut-off at high energies and photoelectric absorption 
at low energies (\cite{Whi83,Nag91}). 
Wheaton et~al.\ (1979)
\nocite{Whe79}, using {\it HEAO1-A1}, first reported the discovery
of an absorption line at $\sim$20~keV. 
Based on {\it
HEAO1-A2} data, White, Swank \& Holt \ (1983) detected cyclotron lines at
$\sim$11.5~keV and $\sim$23~keV, that appeared to be in absorption at the pulse
peak and in emission during the interpulse. By interpreting the two lines 
in terms of the
first and second harmonics of cyclotron resonant scattering, they 
derived $B \simeq 1\times 10^{12}$~G.
During the February 1990 outburst,  
observations with the Large Area Counter onboard {\it Ginga} revelead 
absorption features at $\sim$12~keV and  $\sim$23~keV for all pulse
phases; an investigation of the X--ray spectrum up to 60~keV
did not yield any evidence for higher harmonics 
(\cite{Nag91},\cite{Tam92}). 

On February 22, 1999 the
BATSE instrument onboard the {\it CGRO} satellite revealed the
onset of another outburst of \S\ (\cite{wil99}).
\B\ observed the source with its Narrow Field Instruments (NFI) on four
occasions: 1999 March 6, 19, 22, and 26. The data presented in this {\it
Letter} are from the March 19 observation, when, shortly after the
outburst maximum, the source was at a flux level of $\sim$310~mCrab. 
These data led to the discovery of a four harmonic cyclotron line spectrum in \S,
the first ever from a cosmic X-ray source.  
Our results predate the announcement of the discovery of the 
third cyclotron line in the spectrum of X0115+63 
based on {\it RossiXTE}/HEXTE measurements
(Heindl et al. 1999); therefore they provide also an important independent 
confirmation of the latter result. 

\section{Observations and Spectral Analysis}

Besides the Low-Energy Concentrator
Spectrometer (LECS, 0.1--10~keV, \cite{Par97}) and the Medium-Energy Concentrators
Spectrometer (MECS, 2--10~keV, \cite{Boe97b}), 
the NFIs onboard the \B\ satellite (\cite{Boe97a}) comprise two  
collimated high energy detectors, the High Pressure
Gas Scintillation  Proportional Counter (HPGSPC, 4--60~keV, 
FWHM energy resolution of 8\% at 10 ~keV and 5.5\% at 20~keV, 
\cite{Manz97}),
and the  Phoswich Detection
System (PDS, 15-200~keV, FWHM energy
resolution of 24\% at 20~keV,  
and 14\% at 60~keV, \cite{Fro97}). 

\S\ was observed with the NFIs aboard \B\ from March 19, UT 17:05:25 to March 20, 
08:42:04.
All instruments were operated in their standard configuration. The 
effective exposure
was 3.2~ks for the LECS, 31.4~ks for MECS, 30~ks for HPGSPC and
16~ks for PDS, which makes use of the rocking collimator technique to monitor the 
background. 
The 10--50~keV source flux was $\simeq 1.3\times
10^{-8}$ erg~cm$^{-2}$~s$^{-1}$, corresponding to a luminosity of
$L_{10-50}\simeq2.5\times 10^{37}\ {\rm d}_{4}^2$ erg~s$^{-1}$, where d$_{4}$
is the distance in units of 4~kpc (\cite{Tam92}). 
The source did not show any significant variability during the observation.
In Fig.~\ref{fig:pulses} the pulse profiles folded over the best period of 
$P=3.6144(1)$~s in six different energy bands are reported.
The pulse profile shows the typical double peaked structure, 
already apparent in the {\it HEAO1}
and {\it Ginga} data (\cite{Whi83}; \cite{Nag91}): a 
pronounced main peak (phase
0--0.35), followed by a broader and much softer second peak (phase 0.5--0.85). 
The shape of both peaks is clearly energy-dependent. 

This {\it Letter} concentrates on the high energy X-ray spectrum 
($\sim 9-100$~keV), based on the HPGSPC and PDS data.
In consideration of the 
strong phase
dependence of the cyclotron lines of \S\ (\cite{Miha95}), we accumulated the 
PHA spectra over 10 pulse phase
intervals.
Initially, these spectra
were divided by the PHA spectrum of the Crab Nebula and 
multiplied
by the spectral shape of the Crab Nebula, a power-law with photon 
index, $\alpha$, of 2.1, such that marked spectral features could be spotted 
in an approximately model- and calibration-independent fashion 
(\cite{DAL99}). Dips at 
$\sim$12~keV, $\sim$24~keV, and $\sim$36~keV and, possibly, $\sim$48~keV 
were apparent in the spectra from a number of phase intervals. 
In particular, the spectrum of the descending edge
of the main peak (phase of 0.2--0.3), shown in
Fig.\ref{fig:crab_ratios},  displayed by far the deepest features at
$\sim$12~keV, $\sim$24~keV, $\sim$36~keV and a clear evidence of a dip at
$\sim$48~keV. This motivated us to carry out  
a detailed spectral analysis,
with models including multiple harmonic features.

We used the following 
continuum models to fit the 9--100~keV spectra:
(a) the Negative and Positive power laws EXponential (NPEX) model 
adopted by Mihara
(1995) as the standard model for X--ray Pulsars observed with {\it Ginga}, 
$f(E)=(AE^{-\alpha_1}+BE^{+\alpha_2}) \times \exp(-E/kT)$; (b) a power law with
a high energy cut-off, $f(E)=AE^{-\alpha} \times \exp(-E/kT)$. Here $f(E)$ is 
the photon flux, $kT$ is the e-folding energy and $\alpha$ is the photon index. 
Independent of the continuum model used, at least three
absorption-like features were required in the fit. These 
features were introduced in the model(s) as Gaussian filters in absorption, 
{\it i.e.} $G_i(E)=1-D_i
\times\exp(-(E-E^{\rm cyc}_i)^2/(2\sigma_i^2))$ where $E^{\rm cyc}_i$,
$\sigma_i$, and $D_i$ are the centroid energy, width and depth of each 
feature. Introducing the third absorption feature at $\sim 38$~keV (in addition to 
the first two harmonics at $\sim 12$ and $24$~keV) led to a pronounced 
improvement in the fit, with
the reduced $\chi^2_{\rm dof}$ decreasing from 2.5 (268 dof) to 1.7 (265 dof)
in the case of the NPEX model. 
An F test shows that the probability of chance improvement is of $\sim 
10^{-21}$. 

The HPGSPC and PDS count spectra of the 
descending edge of the main peak (pulse phase 0.2--0.3), together with the
best fit model described above
are shown in Fig.~\ref{fig:spectrum} (upper panel):
An additonal feature centered around $\sim$48~keV is clearly apparent in the residuals of
both the HPGSPC and PDS spectra (Fig.~\ref{fig:spectrum} bottom panel). 
This prompted us to introduce a fourth absorption feature, $G_4$, in the model; 
the minimum $\chi^2_{\rm dof}$
decreased to 1.24 (262 dof, NPEX continuum),
corresponding to an F-test probability of chance improvement of 
$\sim 10^{-15}$. 
Fig.~\ref{fig:residuals} shows the unfolded spectrum of \S. 
Best fit parameters and
equivalent widths are summarized in Table~\ref{tab:best_fit}. In the same Table
best-fit parameters obtained by using the power-law with an exponential 
cut-off are also given. 

We also performed a fit with all the line centroids constrained to an integer harmonic spacing.
The resulting minimum
$\chi^2_{\rm dof}$ is 1.58(259 dof) for the NPEX model and 1.67(259 dof) for
the power law plus cutoff model. An F-test gives a probability of chance
improvement for the models with non-constrained line centroids
$< 10^{-10}$ in both cases.

A preliminary analysis of the spectra from other phase 
intervals also shows significant variations of the line 
features with the pulse phase confirming previous findings from
{\it Ginga} and {\it RossiXTE} (\cite{Heind99}; \cite{Nag91}).
Variations up to 10\% in centroid energy are observed. 
Three lines are still observed 
at the descending edge of the soft broad peak, at phase 0.5--0.6, with centroid
energies of $11.22\pm0.3$~keV, $21.69\pm0.2$~keV and $32.28\pm0.5$~keV.
A comparison with the centroid energies of the three harmonics 
derived by  Heindl et al. (1999), based on {\it RossiXTE} data 
is far from straightforward in consideration of the close but still different 
pulse phase interval (0.7-0.76) over which their spectrum was accumulated
and considering also the different phase of the outburst.

\section{Conclusions}

The spectroscopic capabilities of the high
energy instruments (HPGSPC  and PDS) onboard \B\ allowed to us
study multiple absorption-like features in the spectrum of the 
X-ray pulsar transient \S\ .
In particular four features centered at energies of 
$\sim 12.7$, $24.$,
$36$~keV, and $50$~keV were found in the 
descending edge of the main peak of the pulse profile. 
We fitted the line centroids in Table 1 with a simple linear model 
for the form $E=aN$ (with $N=1,2,3,4$ and $a$ a free parameter). 
Unacceptable values of $\chi^2_{dof}$  were obtained: 79.3(3) for the
power law plus cutoff model and 71.4(3) for the NPEX model. 
We conclude that the line centroids reported in Table 1 are not equispaced,
for both continuum models. Stated differently, the corresponding harmonic
ratios  1:($1.9\pm 0.05$):($2.8\pm 0.05$):($3.9\pm0.1$) 
are significantly different from the classical values 1:2:3:4. 
A closer look at
the data reveals that this result can be ascribed entirely to the value of
the centroid of the first harmonic. As an example, in the case of a power law
plus cutoff model, the first harmonic is at 12.79$\pm$0.05 keV, while a fit
to the other three harmonics gives a spacing of 12.02$\pm$0.02 keV. 
Similar results are obtained by using 
the centroid energies obtained from the NPEX continuum model.
We note that, since the first harmonics lies close to the energy interval 
over which the slope of the X-ray spectrum steepens rapidly, 
the determination of its centroid energy 
could be affected by a somewhat inadequate modelling of the continuum.
For example, 
the excess of ``shoulder" photons, below the fundamental, as predicted in
many theoretical calculations (\cite{Ise98}),  may cause the fitted centroid
energy of the fundamental to  appear higher, therefore causing 
the harmonic ratios to be somewhat smaller than expected. 

It is well know that in strong magnetic fields, 
the energy spacing of cyclotron harmonics is altered by 
relativistic effects, {\it i.e.}  
$E_N = m_e c^2
\{ [1+2n(B/B_{crit})\sin^2\theta]^{1/2}-1\} /\sin^2\theta$,  where $m_e$ is
the electron mass, $\theta$ the angle between the  photon propagation angle
and the B-field, and $B_{crit}=4.414\times  10^{13}$~G (see e.g.
\cite{Araya96,Wang}).
This formula, further corrected for the gravitational redshift, was fit to
the centroid energies of the four harmonics observed by {\it BeppoSAX}. 
While formally unacceptable ($\chi^2_{dof}$ 79.9(2) for the power law plus cutoff
model and 35.7(2) for the NPEX model), the best fit obtained in this way 
was somewhat better than the simple linear fit. 

Despite the uncertainties described above, 
we conclude that the centroid energies of the four spectral features 
of \S\ 
are most naturally interpreted in terms of the fundamental, second, 
third and fourth 
harmonics of cyclotron resonant features in a strong magnetic
field, taking into account relativistic corrections. 
This is the first time that four harmonics are observed in
the X-ray spectrum of any cosmic source. 
We find that equivalent widths of the second,
third, and fourth harmonics are larger than that of
the fundamental, confirming and extending previous 
results from {\it Ginga} (\cite{Nag91}). 
Such a trend was predicted by Alexander \&
M{\'e}sz{\'a}ros (1989, 1991), \nocite{alex89,alex91} 
who found that two-photon scattering and two-photon emission
processes 
have a major effect in determining the depth of the second and higher
harmonics relative to the fundamental. 
Detailed calculations show that, while 
the equivalent width of the second harmonic is always
larger than the fundamental, this is not necessarily the case for the 
third (and the fourth) harmonics. In fact the strength of the third harmonic 
depends
strongly on $\theta$ and the optical depth
(\cite{Ise98}). Comparison of our measured spectra with the
ones calculated by Alexander \& M{\'e}sz{\'a}ros (1991)
\nocite{alex91} shows a qualitative agreement.
A systematic study of the \B\ spectra of \S\ as function of 
pulse phase for different mass accretion rates ({\it i.e.} different 
outburst phases) is currently underway and will be published 
elsewhere.

\acknowledgments

The authors wish to thank Milvia Capalbi of the \B\ Scientific Data
Center and the \B\ Mission Director R. C. Butler.

\newpage

\begin{deluxetable}{llll}
\tablewidth{0pt}
\tablecaption{Best-fit spectral parameters\ \tablenotemark{a}
\label{tab:best_fit}}
\tablehead{
\multicolumn{2}{l}{Parameter} & \multicolumn{2}{c}{Value} \\
 \cline{3-4} &  & \colhead{NPEX} & \colhead{Cutoff PL}
}
\startdata
$\alpha_1$	  &			& $1.37\pm 0.05$  	& $1.3\pm 0.05$        
\nl
$\alpha_2$	  &			& $0.41\pm 0.05$        & \nodata              
\nl
$kT$              & (keV)               & $11.0\pm 0.05$        &  $17.4\pm 0.5$            
\nl
E$_1^{\rm cyc}$   & (keV)               & $12.74\pm 0.08  $     & $12.78\pm 0.08 
 $    
\nl
$\sigma_1     $   & (keV)               & $1.34 \pm 0.25 $      & $1.52 \pm 0.14 
$     
\nl
$D_1$             &                     & $0.21 \pm 0.04 $      & $0.23 \pm 0.02$      
\nl  
EW$_1        $    &                     & $0.75 \pm 0.04 $      & $0.87 \pm 0.07$            
\nl
E$_2^{\rm cyc}$   & (keV)               & $24.16 \pm 0.07$      & $24.0 \pm 0.07$      
\nl
$\sigma_2     $   & (keV)               & $2.11 \pm 0.18 $      & $1.94 \pm 0.11$      
\nl
$D_2      $       &                     & $0.52 \pm 0.02 $      & $0.50 \pm 0.02 $     
\nl
EW$_2         $   &                     & $2.7  \pm 0.07 $      & $2.4 \pm 0.1$             
\nl
E$_3^{\rm cyc}$   & (keV)               & $35.74\pm 0.35$       & $36.00\pm 0.35$      
\nl
$\sigma_3     $   & (keV)               & $2.53 \pm 0.5$        & $1.98 \pm 0.4$       
\nl
$D_3      $       &                     & $0.46 \pm 0.04$       & $0.43 \pm 
0.04$      
\nl 
EW$_3         $   &                     & $2.8  \pm 0.4 $       & $2.13 \pm 0.3$              
\nl
E$_4^{\rm cyc}$   & (keV)               & $49.5\pm 1.2$         & $49.8\pm 1.4$        
\nl
$\sigma_4     $   & (keV)               & $6.3 \pm 2.3$         & $4.8 \pm 2.0$        
\nl
$D_4      $       &                     & $0.35 \pm 0.06$       & $0.3 \pm 0.06$       
\nl  
EW$_4         $   &                     & $5.2  \pm 1.$         & $3.4 \pm 1.$              
\nl
$\chi^{2}_{\rm dof}$ (dof)  &           & 1.24 (262)             & 1.34 (262)           
\nl
\tablenotetext{a}{ Uncertainties at 90\% confidence level for a single 
parameter.}
\enddata
\end{deluxetable}

\ifPreprintMode\relax\else

\newpage

\figcaption[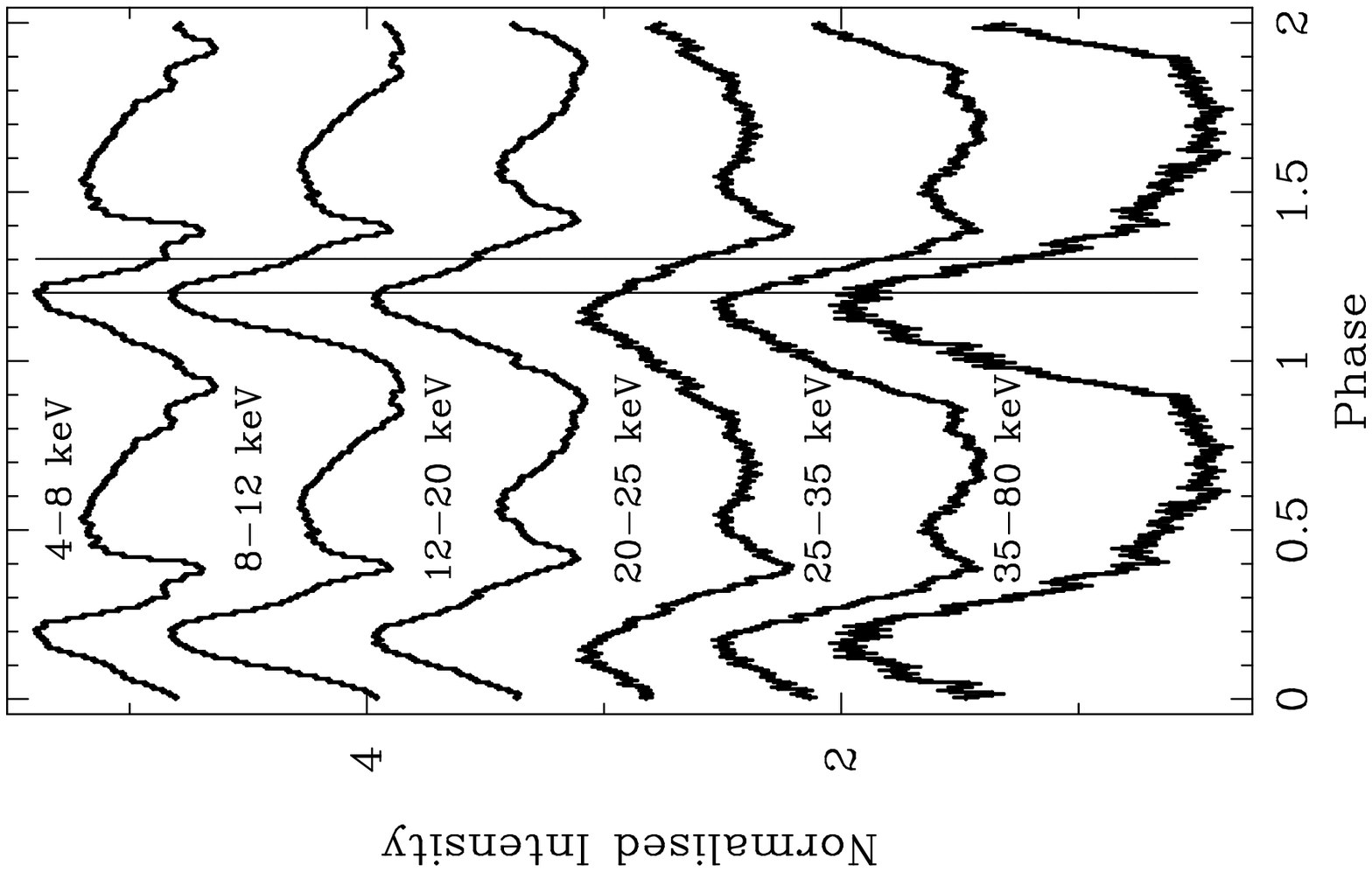]{Pulse profiles of \S\ in six energy bands.
The profiles are normalised to unit average and have been shifted 
vertically by 0.8 for display purposes. 
Data below (above) 20~keV are from the HPGSPC (PDS). 
The vertical lines mark the phase interval over which the 
spectrum of the descending edge of the main pulse has been accumulated. 
The energy dependence of both the main and soft peaks is apparent.
\label{fig:pulses}}

\figcaption[fig02.vps]{Ratio of the HPGSPC ({\it upper panel}) and 
PDS ({\it lower panel}) observed count rate spectra
with respect to the Crab nebula count rate spectrum, multiplied by a power-law
with $\alpha = 2.1$. Spectra correspond to
phase 0.2--0.3, the falling edge of the main peak.
\label{fig:crab_ratios}}

\figcaption[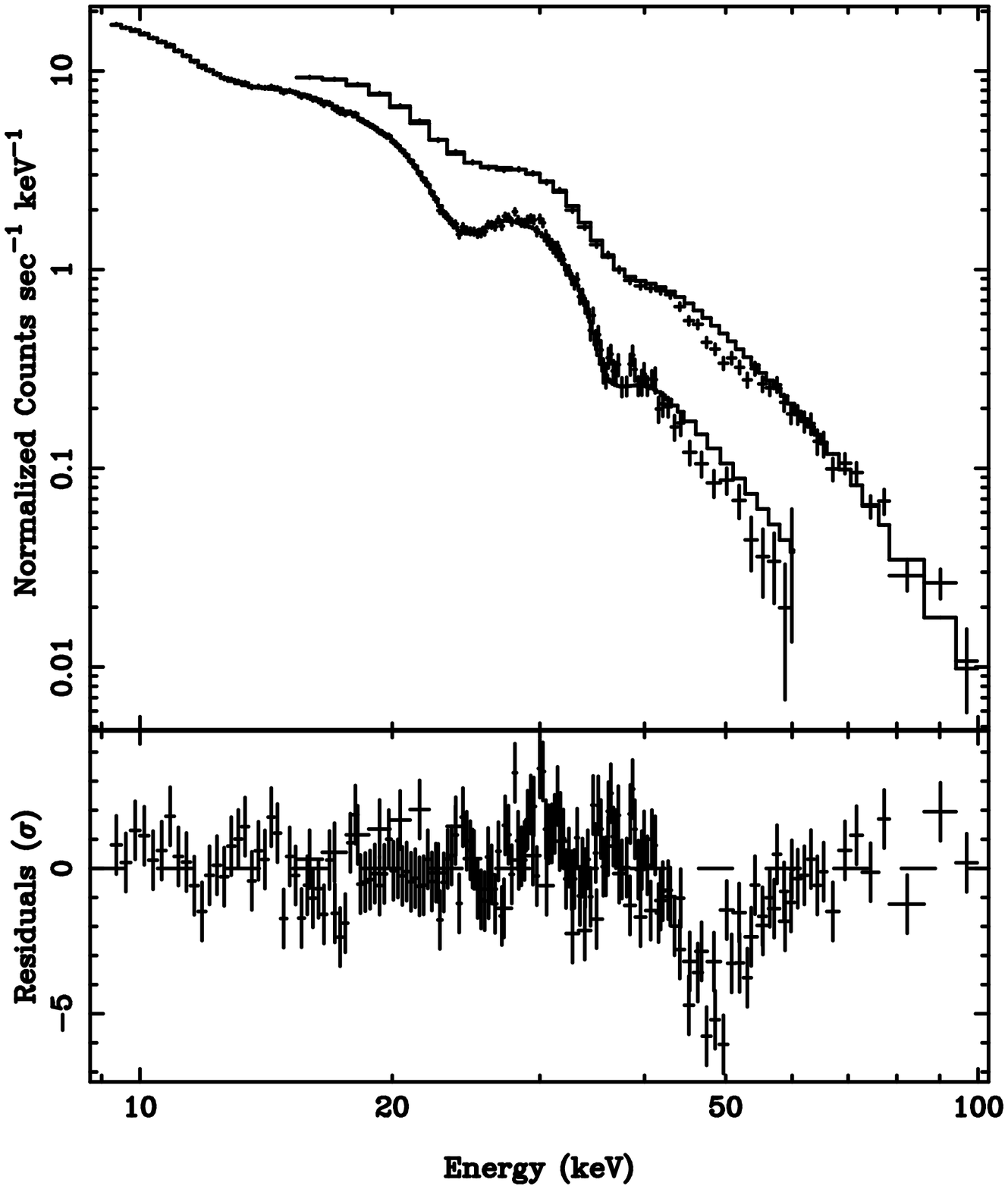]{The 9--100~keV spectrum of \S\ observed by
\B\ in the descending edge of the main peak. Count rate spectra from the
HPGSPC and PDS, together with the best-fit model, which includes the NPEX 
continuum and three absorption features are shown in the upper panel. The
lower panel shows the residuals in units of $\sigma$ revealing evidence for an
absorption feature at $\sim$48~keV.
\label{fig:spectrum}}

\figcaption[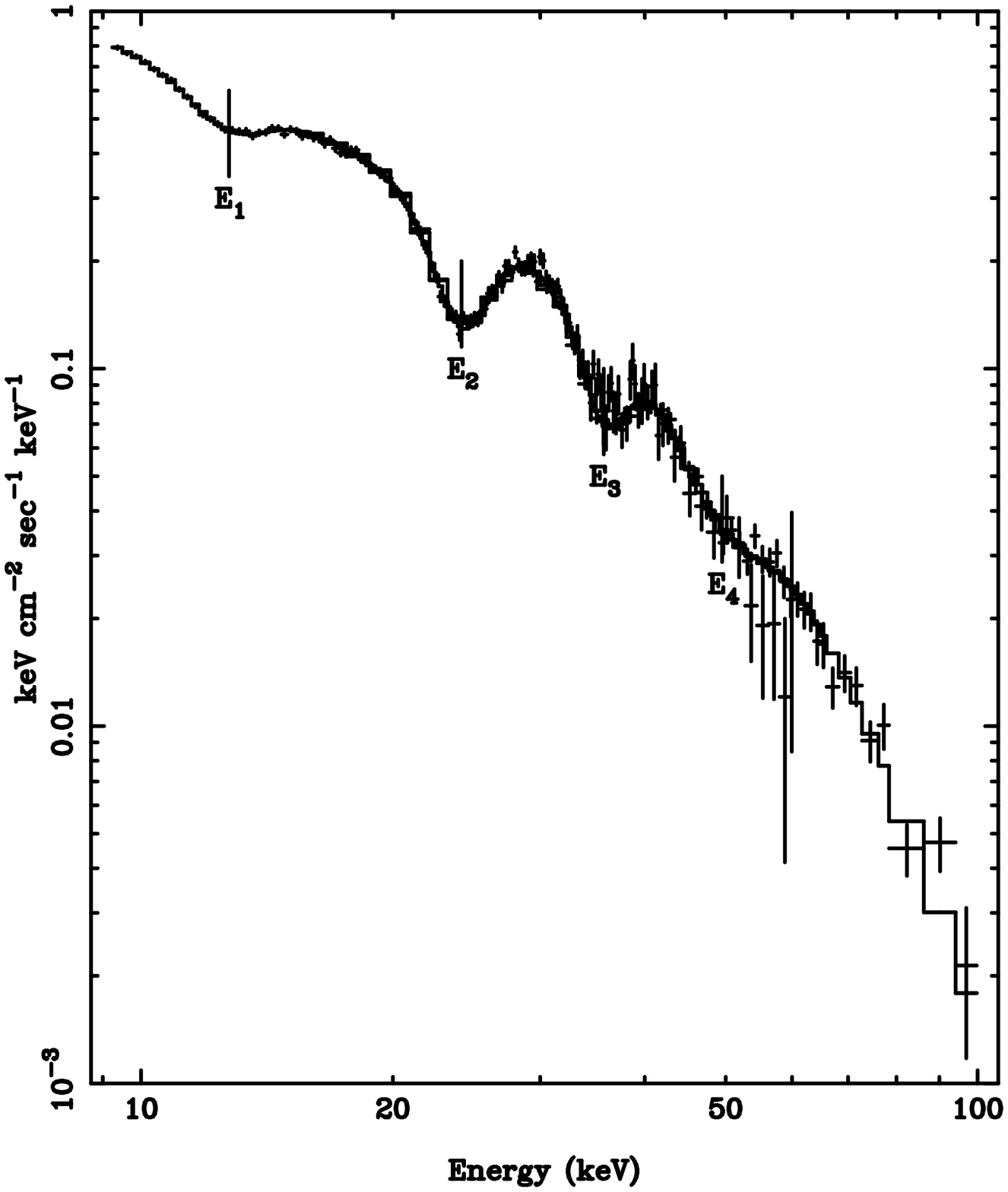]{The unfolded spectrum of the 
descending edge of the main peak of \S\ .
\label{fig:residuals}}

\fi

\newpage

\begin{figure}
\plotone{fig01.ps}
\ifPreprintMode
\caption[]{Pulse profiles of \S\ in six energy bands.
The profiles are normalised to unit average and have been shifted 
vertically by 0.8 for display purposes. 
Data below (above) 20~keV are from the HPGSPC (PDS). 
The vertical lines mark the phase interval over which the 
spectrum of the descending edge of the main pulse has been accumulated. 
The energy dependence of both the main and soft peaks is apparent.}
\label{fig:pulses}
\fi
\end{figure}

\begin{figure}
\plotone{fig02.vps}
\ifPreprintMode
\caption[]{Ratio of the HPGSPC ({\it upper panel}) and 
PDS ({\it lower panel}) observed count rate spectra
with respect to the Crab nebula count rate spectrum, multiplied by a power-law
with $\alpha = 2.1$. Spectra correspond to
phase 0.2--0.3, the falling edge of the main peak.}
\label{fig:crab_ratios}
\fi
\end{figure}

\begin{figure}
\plotone{fig03.ps}
\ifPreprintMode
\caption[]{The 9--100~keV spectrum of \S\ observed by
\B\ in the descending edge of the main peak. Count rate spectra from the
HPGSPC and PDS, together with the best-fit model, which includes the NPEX
continuum and three absorption features are shown in the upper panel. The
lower panel shows the residuals in units of $\sigma$ revealing evidence for an
absorption feature at $\sim$48~keV.}
\label{fig:spectrum}
\fi
\end{figure}

\begin{figure}
\plotone{fig04.ps}
\ifPreprintMode
\caption[]{The unfolded spectrum of \S, corresponding to the falling edge of the 
main peak.}
\label{fig:residuals}
\fi
\end{figure}

\end{document}